\documentclass[english,aip,jcp,reprint]{revtex4-2}
\usepackage[T1]{fontenc}
\usepackage[utf8]{inputenc}
\usepackage{babel}
\usepackage{refstyle}
\usepackage{amsbsy}
\usepackage{amstext}
\usepackage{graphicx}
\usepackage[unicode=true,pdfusetitle,
 bookmarks=true,bookmarksnumbered=false,bookmarksopen=false,
 breaklinks=false,pdfborder={0 0 1},backref=false,colorlinks=false]
 {hyperref}

\makeatletter

\AtBeginDocument{\providecommand\secref[1]{\ref{sec:#1}}}
\AtBeginDocument{\providecommand\eqref[1]{\ref{eq:#1}}}
\AtBeginDocument{\providecommand\schemeref[1]{\ref{scheme:#1}}}
\AtBeginDocument{\providecommand\subsecref[1]{\ref{subsec:#1}}}
\AtBeginDocument{\providecommand\figref[1]{\ref{fig:#1}}}
\RS@ifundefined{subsecref}
  {\newref{subsec}{name = \RSsectxt}}
  {}
\RS@ifundefined{thmref}
  {\def\RSthmtxt{theorem~}\newref{thm}{name = \RSthmtxt}}
  {}
\RS@ifundefined{lemref}
  {\def\RSlemtxt{lemma~}\newref{lem}{name = \RSlemtxt}}
  {}

\usepackage[version=4]{mhchem}
\usepackage{notes2bib}

\bibnotesetup{
note-name = ,
use-sort-key = false
}

\makeatother

\begin{document}
\title{Many recent density functionals are numerically ill-behaved}
\author{Susi Lehtola}
\email{susi.lehtola@alumni.helsinki.fi}

\affiliation{Molecular Sciences Software Institute, Blacksburg, Virginia 24061,
United States}
\address{Department of Chemistry, University of Helsinki, P.O. Box 55, FI-00014
University of Helsinki, Finland}
\author{Miguel A. L. Marques}
\affiliation{Institut für Physik, Martin-Luther-Universität Halle-Wittenberg, 06120
Halle (Saale), Germany}
\begin{abstract}
Most computational studies in chemistry and materials science are
based on the use of density functional theory. Although the exact
density functional is unknown, several density functional approximations
(DFAs) offer a good balance of affordable computational cost and semi-quantitative
accuracy for applications. The development of DFAs still continues
on many fronts, and several new DFAs aiming for improved accuracy
are published every year. However, the numerical behavior of these
DFAs is an often overlooked problem. In this work, we look at all
592 DFAs for three-dimensional systems available in Libxc 5.2.2 and
examine the convergence of the density functional total energy based
on tabulated atomic Hartree--Fock wave functions. We show that several
recent DFAs, including the celebrated SCAN family of functionals,
show impractically slow convergence with typically used numerical
quadrature schemes, making these functionals unsuitable both for routine
applications or high-precision studies, as thousands of radial quadrature
points may be required to achieve sub-$\mu E_{h}$ accurate total
energies for these functionals, while standard quadrature grids like
the SG-3 grid only contain $\mathcal{O}(100)$ radial quadrature points.
These results are both a warning to users to always check the sufficiency
of the quadrature grid when adopting novel functionals, as well as
a guideline to the theory community to develop better behaved density
functionals.
\end{abstract}
\maketitle
\newcommand*\ie{{\em i.e.}}
\newcommand*\eg{{\em e.g.}}
\newcommand*\etal{{\em et al.}}
\newcommand*\citeref[1]{ref. \citenum{#1}}
\newcommand*\citerefs[1]{refs. \citenum{#1}} 

\newcommand*\Erkale{{\sc Erkale}}
\newcommand*\Bagel{{\sc Bagel}}
\newcommand*\FHIaims{{\sc FHI-aims}}
\newcommand*\LibXC{{\sc LibXC}}
\newcommand*\Orca{{\sc Orca}}
\newcommand*\PySCF{{\sc PySCF}}
\newcommand*\PsiFour{{\sc Psi4}}
\newcommand*\Turbomole{{\sc Turbomole}}

\let\schemeref\undefined
\newref{scheme}{name=scheme~, names=schemes~, Name=Scheme~, Names=Schemes~}

\section{Introduction \label{sec:Introduction}}

Computational studies in chemistry and materials science are typically
based on the use of density functional theory\citep{Hohenberg1964_PR_864,Kohn1965_PR_1133}
(DFT).\citep{Barth2004_PS_9,Becke2014_JCP_18,Kryachko2014_PR_123,Jones2015_RMP_897,Mardirossian2017_MP_2315}
The reason for the usefulness of this theory is that the complicated
quantum mechanical interactions between the electrons can be reduced
to consideration of the electron density, only, thus making calculations
much simpler and more affordable than those with traditional wave
function methods.\citep{Kohn1999_RMP_1253}

Fully numerical methods\citep{Lehtola2019_IJQC_25968} have recently
enabled reliable computations of DFT total energies for moderate sized
systems to sub-microhartree accuracy, that is, at the complete basis
set (CBS) limit.\citep{Jensen2016_PCCP_21145,Jensen2017_JPCL_1449,Lehtola2019_IJQC_25944,Lehtola2019_IJQC_25945,Lehtola2020_PRA_12516,Brakestad2020_JCTC_4874,Brakestad2021_JCP_214302}
However, in order for fully numerical calculations to be tractable,
the density functional approximations (DFAs) used in the calculations
have to be well-behaved. Our (S.L.) recent results on the determination
of total atomic energies at the CBS limit with fully numerical methods
with meta-GGA functionals along the lines of \citerefs{Lehtola2019_IJQC_25945}
and \citenum{Lehtola2020_PRA_12516} suggest that many functionals---including
recent ones---are problematic in this aspect.

Determining accurate total energies with fully numerical methods\citep{Lehtola2019_IJQC_25968}
requires being able to run self-consistent field (SCF) calculations\citep{Lehtola2020_M_1218}
in extended basis sets that approach the CBS limit. While the SCF
procedure can be carried out using various techniques, such as Roothaan's
method of iterative diagonalization\citep{Roothaan1951_RMP_69} or
orbital rotation techniques,\citep{HeadGordon1988_JPC_3063} for example,
regardless of the employed approach, the determination of reliable
CBS limit total energies requires the ability to
\begin{enumerate}
\item evaluate the total energy accurately for a fixed electron density,
as in an individual SCF step, \label{enu:accurate-evaluation-of}
\item converge the iterative SCF procedure tightly in a given one-particle
basis set, yielding an optimized density and total energy in the fixed
basis, and \label{enu:tight-convergeability-of-SCF}
\item smoothly converge the total energy to the CBS limit by running SCF
calculations in a systematic sequence of larger and larger one-particle
basis sets, eventually reaching a value converged to sub-$\mu E_{h}$
precision.\citep{Lehtola2019_IJQC_25945,Lehtola2020_PRA_12516} \label{enu:cbs-limit}
\end{enumerate}
The above three criteria can be used to study density functional approximations.
It is already known that many functionals fail the latter two requirements.
For instance, the local $\tau$ approximation of \citet{Ernzerhof1999_JCP_911}
and related functionals\citep{Lehtola2021_JCTC_943} produce diverging
potentials,\citep{Lehtola2021_JCTC_943} which complicates SCF calculations
even in small basis sets, thereby breaking criterion \ref{enu:tight-convergeability-of-SCF}.
Next, many Minnesota functionals are known to exhibit pathologically
slow convergence to the CBS limit,\citep{Mardirossian2013_JCTC_4453}
breaking criterion \ref{enu:cbs-limit}. For instance, unexpectedly
large $mE_{h}$ level basis set truncation errors in standard quadruple-$\zeta$
Gaussian basis sets in their uncontracted form were recently observed
for the M11-L functional\citep{Peverati2012_JPCL_117} already for
hydrogen, while the truncation errors for well-behaved functionals
were found to be around two orders of magnitude smaller.\citep{Schwalbe2022__}

We will show in this work that many recent functionals fail already
for criterion \ref{enu:accurate-evaluation-of}, which precedes any
SCF calculation. Using tabulated Hartree--Fock wave functions for
atoms, we demonstrate that the total energies obtained with standard
sized quadrature grids are unreliable. Although reliable total energies
can be obtained in principle by using uncustomarily large quadrature
grids, we show that several functionals like the celebrated SCAN family\citep{Sun2015_PRL_36402,Bartok2019_JCP_161101,Furness2020_JPCL_8208,Furness2020_JPCL_9248,Furness2022_JCP_34109}
require impractically many radial quadrature points (thousands instead
of around one hundred) to converge to the level of accuracy expected
in routine applications of quantum chemistry, as the default SCF convergence
settings in most programs require evaluating total energies to sub-$\mu E_{h}$
precision. As polyatomic systems like molecules and crystals are made
from atoms, the issues found in this work also have ramifications
to practical applications of these density functionals, suggesting
that more work is needed to develop accurate functionals that satisfy
all of the three criteria given above.

The layout of this work is as follows. A brief summary of DFT is given
in \secref{Theory}. The computational details of are presented in
\secref{Computational-Details}, while the results of our approach
are discussed in \secref{Results}. A summary is presented in \secref{Summary-and-Discussion},
followed by discussion. Atomic units are used throughout unless specified
otherwise.

\section{Theory \label{sec:Theory}}

In DFT, the total energy is expressed as
\begin{equation}
E[n_{\uparrow},n_{\downarrow}]=T[n_{\uparrow},n_{\downarrow}]+V[n]+E_{J}[n]+E_{\text{xc}}[n_{\uparrow},n_{\downarrow}],\label{eq:Etot}
\end{equation}
where $n_{\uparrow}$ and $n_{\downarrow}$ are the spin-up and spin-down
electron densities and $n=n_{\uparrow}+n_{\downarrow}$ is the electron
density, $T$ is the kinetic energy (typically evaluated in terms
of the occupied orbitals as suggested by \citet{Kohn1965_PR_1133}),
$V$ is the nuclear attraction energy, $E_{J}$ is the classical Coulomb
repulsion of the electrons, and $E_{\text{xc}}$ is the quantum mechanical
exchange-correlation energy. Common DFAs express $E_{\text{xc}}$
as 

\begin{equation}
E_{\text{xc}}[n]=\int n\epsilon_{\text{xc}}(n_{\uparrow},n_{\downarrow},\gamma_{\uparrow\uparrow},\gamma_{\uparrow\downarrow},\gamma_{\downarrow\downarrow},\nabla^{2}n_{\uparrow},\nabla^{2}n_{\downarrow},\tau_{\uparrow},\tau_{\downarrow}){\rm d}^{3}r,\label{eq:Edft}
\end{equation}
where $\gamma_{\sigma\sigma'}$ are reduced gradients
\begin{equation}
\gamma_{\sigma\sigma'}=\nabla n_{\sigma}\cdot\nabla n_{\sigma'}\label{eq:redgrad}
\end{equation}
and $\tau_{\uparrow}$ and $\tau_{\downarrow}$ are the local kinetic
energy densities
\begin{equation}
\tau_{\sigma}=\frac{1}{2}\sum_{i\text{ occupied}}|\nabla\psi_{i\sigma}|^{2},\label{eq:tau}
\end{equation}
and $\sigma$ and $\sigma'$ are a spin indices. The $\epsilon_{\text{xc}}$
term in \eqref{Edft} is the DFA, which is a (often complicated) mathematical
function with known analytical form. DFAs can be classified on Jacob's
ladder\citep{Perdew2001_ACP_1} based on their ingredients:
\begin{itemize}
\item local density approximation (LDA): dependence only on the local electron
density $n_{\uparrow}$ and $n_{\downarrow}$
\item meta-LDA approximation:\citep{Lehtola2021_JCTC_943} dependence on
$n_{\uparrow}$ and $n_{\downarrow}$ as well as the local kinetic
energy density $\tau_{\uparrow}$ and $\tau_{\downarrow}$
\item generalized-gradient approximation (GGA): dependence on $n_{\uparrow}$
and $n_{\downarrow}$ as well as their gradients $\nabla n_{\uparrow}$
and $\nabla n_{\downarrow}$ through $\gamma_{\sigma\sigma'}$
\item meta-GGA approximation: further dependence on the Laplacian $\nabla^{2}n_{\uparrow}$,
$\nabla^{2}n_{\downarrow}$, and/or $\tau_{\uparrow}$, $\tau_{\downarrow}$
\end{itemize}
In addition to a term of the form of \eqref{Edft}, many common DFAs
also add post-DFT terms such as 
\begin{itemize}
\item exact exchange in either the full Hartree--Fock (global hybrids,
e.g. the B3LYP functional\citep{Stephens1994_JPC_11623}) or range-separated\citep{Gill1996_MP_1005,Leininger1997_CPL_151}
form (range-separated hybrids, e.g. the $\omega$B97X functional\citep{Chai2008_JCP_84106})
\item non-local correlation (e.g. $\omega$B97X-V\citep{Mardirossian2014_PCCP_9904})
or semiempirical dispersion (e.g. the $\omega$B97X-D3 functional\citep{Lin2013_JCTC_263})
\item post-Hartree--Fock correlation (double hybrids, e.g. the XYG3 functional\citep{Zhang2009_PNAS_4963})
\end{itemize}
These additional ingredients will not be discussed further in this
work as they are not thought to present major issues with numerical
behavior. Instead, the issues with numerical ill behavior arise mainly
from \eqref{Edft}. 

For completeness, we note here that local hybrid functionals, which
include a position-dependent DFA-type fraction of exact exchange energy
density, have also been suggested.\citep{Jaramillo2003_JCP_1068,Maier2019_WIRCMS_1378}
However, as i) there are fewer local hybrids than functionals that
fit in the above classification, ii) local hybrids have not become
widely used, and iii) the analysis of numerical ill behavior in local
hybrids is not as straightforward to study as that arising from \eqref{Edft},
we do not consider local hybrid functionals in this work.

Programs that employ atomic orbital basis sets typically evaluate
\eqref{Edft} using the multicenter quadrature approach developed
by \citet{Becke1988_JCP_2547}. By inserting a resolution of the identity
\begin{equation}
\sum_{A}w_{A}(\boldsymbol{r})=1\label{eq:atomic-weight}
\end{equation}
 in terms of atomic weight functions $w_{A}(\boldsymbol{r})$,\citep{Becke1988_JCP_2547,Stratmann1996_CPL_213,Laqua2018_JCP_204111}
the integral in \eqref{Edft} can be evaluated as a sum of atom-centered
integrals 
\begin{equation}
\int f(\boldsymbol{r}){\rm d}^{3}r=\sum_{A}\int_{A}f(\boldsymbol{r})w_{A}(\boldsymbol{r}){\rm d}^{3}r.\label{eq:atint}
\end{equation}
The atom-centered integrals are evaluated on a grid obtained as the
tensor product of a radial quadrature grid\citep{Murray1993_MP_997,Treutler1995_JCP_346,Mura1996_JCP_9848,Krack1998_JCP_3226,Lindh2001_TCA_178,Gill2003_JCC_40}
\begin{equation}
\int_{0}^{\infty}r^{2}f(r){\rm d}r\approx\sum_{p}w_{p}f(r_{p})\label{eq:radrule}
\end{equation}
and an angular grid
\begin{equation}
\int f(\boldsymbol{\Omega}){\rm d}\Omega\approx\sum_{q}w_{q}f(\Omega_{q})\label{eq:angrule}
\end{equation}
which is almost invariably a Lebedev grid\citep{Lebedev1975_UCMMP_44,Lebedev1976_UCMMP_10,Lebedev1977_SMJ_99,Lebedev1992_RASDM_587,Lebedev1995_RASDM_283},
although other types of grids have also been suggested.\citep{Murray1993_MP_997,Daul1997_IJQC_219}
Note that programs that do not employ atomic-orbital basis sets also
use quadrature to evaluate \eqref{Edft}, meaning that any ill behavior
found in this work may have ramifications to also such othe approaches.

Regardless of the approach used, the quadrature error in \eqref{Edft}
can be made negligible by using sufficiently many points. Our hypothesis
is that the convergence of the quadrature with the number of points
is intimately related to the numerical well-behavedness of the DFA.
This leads to the question if the calculation converges quickly enough
for the DFAs that are currently available to allow the determination
of total energies with high precision. 

\section{Computational Details \label{sec:Computational-Details}}

\subsection{Radial quadrature \label{subsec:Radial-quadrature}}

Our main focus is the study of the numerical well-behavedness of \eqref{Edft}
based on tabulated atomic Hartree--Fock wave functions. In the case
of a single atom, the Becke weighting yields an unit weight. The electron
density arising from the Hartree--Fock wave function is assumed to
be spherically symmetric, $n_{\sigma}(\boldsymbol{r})=n_{\sigma}(r)$,
and the integral in \eqref{Edft} thus reduces to a radial one
\begin{equation}
E_{\text{xc}}[n]=4\pi\int_{0}^{\infty}r^{2}n\epsilon_{\text{xc}}(n_{\uparrow},n_{\downarrow},\dots){\rm d}r.\label{eq:Erad}
\end{equation}
We evaluate this integral by $N$-point quadrature 
\begin{equation}
E_{\text{xc}}(N)=4\pi\sum_{i=1}^{N}w_{i}r_{i}^{2}n_{i}\epsilon_{\text{xc}}(n_{i;\uparrow},n_{i;\downarrow},\dots).\label{eq:Equad}
\end{equation}
Several kinds of radial quadratures are considered. Each type of quadrature
is expressed as a coordinate transformation $r=r(x)$ from a primitive
quadrature coordinate $x$. The primitive quadrature can be over $x\in[-1,1]$
or $x\in[0,1]$, depending on the rule. 

Note that most rules were originally developed in combination with
atomic size adjustments, $r=r(x)\to r=Rr(x)$, where $R$ is a element
specific parameter, as in the original scheme of \citet{Becke1988_JCP_2547}.
The atomic scaling changes the radii linearly $r_{i}\to Rr_{i}$ and
the weights cubically $w_{i}\to R^{3}w_{i}$. As the quadratures anyway
approach exactness with $N\to\infty$ quadrature points, for simplicity
we do not consider size adjustments in this work and set $R=1$.

The following radial schemes will be considered:
\begin{enumerate}
\item The M3 grid of \citet{Treutler1995_JCP_346}
\begin{equation}
r=\frac{1}{\ln2}\ln\frac{2}{1-x}\label{eq:radial-transform}
\end{equation}
with $x\in[-1,1]$. \label{scheme:Ahlrichs-M3}
\item The M4 grid of \citet{Treutler1995_JCP_346}, again with the atomic
scaling parameter set to $\xi=1$
\begin{equation}
r=\frac{1}{\ln2}(1+x)^{\alpha}\ln\frac{2}{1-x}\label{eq:ahlrichs-m4}
\end{equation}
with $x\in[-1,1]$ and where $\alpha=0.6$ is the optimized value
of \citet{Treutler1995_JCP_346}; the case $\alpha=0$ reduces to
the M3 quadrature of \eqref{radial-transform}. \label{scheme:Ahlrichs-M4}
\item The scheme of \citet{Murray1993_MP_997} is given by
\begin{equation}
r=\left(\frac{x}{1-x}\right)^{2}\label{eq:euler}
\end{equation}
with $x\in[0,1]$; this transform was originally introduced by \citet{Handy1973_TCA_195}.
This scheme is commonly referred to as Euler--Maclaurin quadrature.
\label{scheme:Euler-Maclaurin}
\item The \citet{Mura1996_JCP_9848} scheme
\begin{equation}
r=-\log(1-x^{m})\label{eq:mura}
\end{equation}
with $x\in[0,1]$ and $m=3$ which is the recommended value for molecular
systems. \label{scheme:Mura-Knowles}
\end{enumerate}
These schemes were chosen to be representative of quantum chemistry
programs in general: by default, \schemeref{Ahlrichs-M3} is used
in ERKALE\citep{Lehtola2012_JCC_1572} and ORCA;\citep{Neese2020_JCP_224108}
\schemeref{Ahlrichs-M4} is used in TURBOMOLE\citep{Balasubramani2020_JCP_184107},
Psi4,\citep{Smith2020_JCP_184108} and PySCF\citep{Sun2020_JCP_24109};
\schemeref{Euler-Maclaurin} is used in Gaussian\bibnotemark[Scalmani] \bibnotetext[Scalmani]{Giovanni Scalmani, Gaussian Inc, private communication, 2022.}
and Q-Chem;\citep{Epifanovsky2021_JCP_84801} and \schemeref{Mura-Knowles}
is used in Molpro\citep{Werner2020_JCP_144107} and NWChem,\citep{Apra2020_JCP_184102}
for example. Other types of radial grids have also been proposed,\citep{Krack1998_JCP_3226,Lindh2001_TCA_178,Kakhiani2009_CPC_256,Mitani2011_TCA_645,Mitani2012_TCA_1169,Graefenstein2007_JCP_164113,ElSherbiny2004_JCC_84,Weber2004_CPC_133,Shizgal2016_JMC_413}
but the four schemes above suffice for the present purposes.

Gauss--Chebyshev quadrature over $x$ was used in the pioneering
work by \citet{Becke1988_JCP_2547}, who preferred this quadrature
over others thanks to its quadrature points and weights being given
by simple analytical formulas. Gauss--Chebyshev quadrature was likewise
used by \citet{Treutler1995_JCP_346}. However, equations for the
quadrature rules were not provided in \citerefs{Becke1988_JCP_2547}
and \citenum{Treutler1995_JCP_346}. In contrast, the scheme of \citet{Murray1993_MP_997}
relies on the special properties\citep{Handy1973_TCA_195} of \eqref{euler}
in combination with the Euler--Maclaurin quadrature formula to obtain
a simple quadrature rule that coincides with the trapezoidal rule.\citep{Gill2003_JCC_40} 

In this work, the primitive $N$-node quadratures in $x$ are generated
with closed-form Gauss--Chebyshev quadrature formulas of the second
kind given by \citet{PerezJorda1994_JCP_6520}, see their equations
(31)--(33) {[}N.B. this is not the rule used for radial quadrature
in \citeref{PerezJorda1994_JCP_6520}!{]}, unless specified otherwise.
The weights are derived from Gauss--Chebyshev quadrature formulas,
$\int_{-1}^{1}f(x)\sqrt{1-x^{2}}{\rm d}x=\sum_{i=1}^{N}w_{i}f(x_{i})$
with $x_{i}=\cos[i\pi/(N+1)]$ and $w_{i}=\pi\sin^{2}[i\pi/(N+1)]/(N+1)$
with a change of variables that converts the rule to the form with
a unit weight function, $\int_{-1}^{1}f(x){\rm d}x=\sum_{i=1}^{N}w_{i}f(x_{i})$.\citep{PerezJorda1992_CPC_271}
The quadrature for $x\in[0,1]$ for \schemeref{Euler-Maclaurin, Mura-Knowles}
is obtained by the change of variables $x\to x'=(1+x)/2$.

For comparison, we also examine trapezoidal quadrature for \schemeref{Euler-Maclaurin, Mura-Knowles}.
\Schemeref{Euler-Maclaurin} was originally described with such a
rule, and \citet{Gill2003_JCC_40} describe \schemeref{Mura-Knowles}
to likewise use trapezoidal quadrature. Trapezoidal nodes and weights
for $x\in[0,1]$ given by $x_{i}=i/(N+1)$ and $w_{i}=1/(N+1)$ were
used with $i\in[1,N]$;\citep{Gill2003_JCC_40} the corresponding
rule for $x\in[-1,-1]$ is $x_{i}=1-2i/(N+1)$ with $w_{i}=2/(N+1)$.

We examine the numerical well-behavedness of DFAs by studying the
convergence of the energy given by \eqref{Erad} with the number of
quadrature points within each radial scheme. As was already mentioned
in \secref{Introduction}, the motivation for this approach is that
it is widely used in DFT calculations for molecular and solid state
systems, and any results from this work therefore immediately generalize
into a larger context. For example, electronic structure calculations
are often started from either atomic densities\citep{Almloef1982_JCC_385,VanLenthe2006_JCC_32}
or atomic potentials\citep{Lehtola2019_JCTC_1593}; the use of tabulated
Hartree--Fock densities corresponds to the former approach and quadrature
errors in the total energy seen for gas-phase atoms will also be observable
for the superposition of atomic densities in polyatomic calculations.

\Eqref{Equad} should show robust convergence for well-behaved DFAs.
This means that it should be possible to bind the quadrature error
to be smaller than a preset threshold $\epsilon$ as 
\begin{equation}
|E_{\text{xc}}(N)-E_{\text{xc}}^{\text{ref}}|\leq\epsilon\label{eq:boundedness}
\end{equation}
for all $N\geq N'$ given some suitable choice of $N'$ (which depends
on the chosen value for $\epsilon$), with the refence value $E_{\text{xc}}^{\text{ref}}$
in \eqref{boundedness} being given by the exact value of the integral
that should be obtainable at the limit 
\begin{equation}
E_{\text{xc}}^{\text{ref}}=\lim_{N\to\infty}E_{\text{xc}}(N).\label{eq:Elimit}
\end{equation}
It is easy to see from \eqref{boundedness} that the difference 
\begin{equation}
\delta(N_{1},N_{2})=|E_{\text{xc}}(N_{1})-E{}_{\text{xc}}(N_{2})|\label{eq:Ediff}
\end{equation}
should also be bounded by an arbitrarily small value for sufficiently
large $N_{1}$ and $N_{2}$, that is, whenever $N_{1},N_{2}\geq N'$.
However, we will demonstrate later in this work that the measure $\delta$
is not small for many recent DFAs.

\subsection{Atomic wave functions \label{subsec:Atomic-wave-functions}}

The best known example of tabulated Hartree--Fock wave functions
is the seminal work of \citet{Clementi1974_ADNDT_177}. However, the
Clementi--Roetti wave functions are inconvenient, because the tables
are not in machine readable format and thus require error-prone parsing.
Moreover, the Clementi--Roetti wave functions have limited accuracy---they
are given at fixed precision with five decimals---and are deficient
for heavy atoms: \citet{Koga1993_PRA_4510} reported reoptimized wave
functions with energies improved by as much as $53\ \text{m}E_{h}$
(for Cd), while \citet{Koga1993_JPBAMOP_2529} found few-$\text{m}E_{h}$
improvements over Clementi and Roetti for all cations and anions with
$Z\ge37$ as well as a staggering $1.8\ E_{h}$ decrease for \ce{Tc-}. 

Better-quality wave functions have been since published. \citet{Koga1999_IJQC_491}
reported wave functions for light elements with $\mu E_{h}$ truncation
errors over numerical Hartree--Fock (NHF) calculations, while \citet{Koga2000_TCA_411}
reported wave functions for heavy elements with $\text{m}E_{h}$ level
truncation errors compared to NHF. Although the Hartree--Fock wave
functions of \citet{Koga1999_IJQC_491} and \citet{Koga2000_TCA_411}
are also reported at fixed precision, the tabulation including 7 decimals,
the wave functions are sufficient for the purposes of this work. 

The wave functions of \citet{Koga1999_IJQC_491} and \citet{Koga2000_TCA_411}
are available \bibnotemark[Thakkar] \bibnotetext[Thakkar]{Ajit Thakkar, private communication, 2020.}
in a simple and easy-to-use Python package called AtomicOrbitals.\citep{Furness__}
AtomicOrbitals allows for easy access to the atomic density data that
can be stored to disk and read into custom implementations of novel
DFAs. As part of this work, AtomicOrbitals was interfaced to Libxc,
and the quadrature approaches discussed in \subsecref{Radial-quadrature}
were implemented therein.

\section{Results \label{sec:Results}}

In our experience, lithium and nitrogen are often especially hard
cases for many functionals, as the electron density ranges from essentially
full spin restriction at the nucleus to full spin polarization far
away, where the electron density is dominated by the slowliest decaying
orbital which is only partially occupied. An analysis of the\textbf{
}592 functionals for three-dimensional systems in Libxc 5.2.2 was
performed for the lithium, nitrogen, neon, sodium, phosphorus, and
argon atoms with the tabulated Hartree--Fock wave functions of \citet{Koga1999_IJQC_491}.
These atoms all have either closed-shell or half-closed $^{n}S$ ground
states, meaning that their ground state naturally has a spherically
symmetric electron density. 

We will base our analysis on the difference (\eqref{Ediff}) of the
$N$-point quadrature formula from the 2501-point quadrature formula
\begin{equation}
|\Delta E|(N)=\delta(N,2501)=|E_{\text{xc}}(N)-E_{\text{xc}}(2501)|.\label{eq:quad-error}
\end{equation}
As discussed in \subsecref{Radial-quadrature}, this measure can be
expected to be reasonable for well-behaved density functionals.

This analysis, performed by visual examination of plots of $\Delta$,
revealed many interesting results. Due to the large number of examined
functionals, figures are shown in the main text only for the key points
of our discussion. We will also mostly limit the discussion to functionals
with numerical ill behavior greater than 1 $nE_{h}$, as energy differences
smaller than this are not thought to cause concern even for high-precision
applications. The full set of figures is available in the Supporting
Information (SI).

\begin{figure}
\begin{centering}
\includegraphics[width=0.5\textwidth]{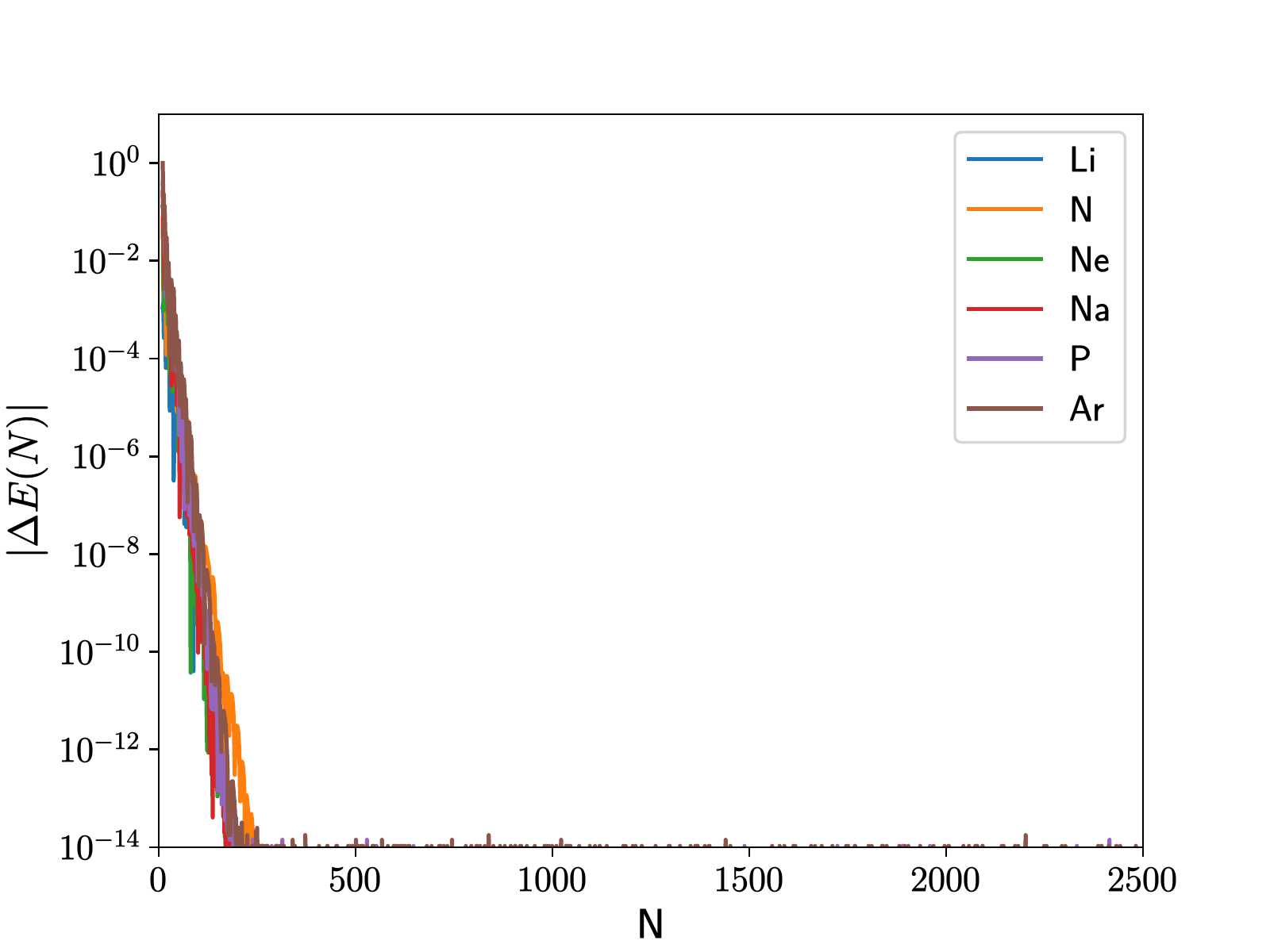}
\par\end{centering}
\caption{Quadrature error (\eqref{quad-error}) for the TPSS meta-GGA exchange
functional as a function of the number of radial points. \label{fig:TPSS-exchange}}
\end{figure}

Our first and main finding is that ``well-behaved'' DFAs exhibit fast
convergence. For example, the LDA,\citep{Bloch1929_ZfuP_545,Dirac1930_MPCPS_376}
the Perdew--Burke--Ernzerhof\citep{Perdew1996_PRL_3865,Perdew1997_PRL_1396}
(PBE) GGA, the Tao--Perdew--Scuseria--Staroverov\citep{Tao2003_PRL_146401,Perdew2004_JCP_6898}
(TPSS) meta-GGA, as well as the recent TASK\citep{Aschebrock2019_PRR_33082}
meta-GGA exchange functionals converge rapidly, only requiring a few
hundred radial quadrature points to converge \eqref{Erad} to better
than $10^{-14}E_{h}$ (which is essentially machine precision) for
the fixed densities. This is illustrated by the TPSS exchange functional
in \figref{TPSS-exchange}. 

The results can be contrasted with those of ``ill-behaved'' functionals,
which we will discuss in the following.  However, we will first simplify
the analysis to a single radial scheme. 

\subsection{Radial quadratures}

As was already argued in \secref{Theory}, the four radial schemes
discussed in \subsecref{Radial-quadrature} are in general found to
yield results of similar quality when Chebyshev quadrature is used:
well-behaved functionals are found to converge to a similar level
of precision with a similar number of quadrature points, while any
pathological behavior is similarly reproducible with any of the studied
radial grids. 

Trapezoidal quadrature was also studied for \schemeref{Euler-Maclaurin, Mura-Knowles}.
While trapezoidal quadrature was found to be competitive for \schemeref{Euler-Maclaurin}
for the case of low numbers of grid points and modest error thresholds
with Chebyshev quadrature, Chebyshev quadrature becomes noticeably
more accurate than trapezoidal quadrature for large numbers of radial
grid points for many well-behaved DFAs. However, there are also many
DFAs for which the opposite conclusion applies. Striking examples
include the CCDF\citep{Margraf2019_JCP_244116} (GGA\_C\_CCDF) and
GAPloc\citep{Fabiano2014_JCTC_2016} (GGA\_C\_GAPLOC) GGA correlation
functionals, which will be further discussed in \subsecref{Ill-behaved-GGAs},
for which \schemeref{Euler-Maclaurin} requires roughly three times
more quadrature points to reach machine precision with Chebyshev quadrature
than with trapezoidal quadrature.

In contrast to the description of the Mura--Knowles scheme (\schemeref{Mura-Knowles})
in \citet{Gill2003_JCC_40}, trapezoidal quadrature was found to yield
extremely poor accuracy for the Mura--Knowles scheme and this combination
will therefore not be considered further in this work.

Overall, the Ahlrichs radial grids (\schemeref{Ahlrichs-M3, Ahlrichs-M4})
appear to afford the best convergence, followed by the Mura--Knowles
scheme with Chebyshev quadrature. The performance of the Euler--Maclaurin
scheme (\schemeref{Euler-Maclaurin}) with either trapezoidal or Chebyshev
weights is found to be less systematic than that of the Ahlrichs grids
or the Mura--Knowles grid. Based on these findings, the figures presented
herein (including the already-shown \figref{TPSS-exchange}) use \schemeref{Ahlrichs-M3}
exclusively, that is, the Ahlrichs M3 grid, in combination with Chebyshev
quadrature. Plots for all studied DFAs with all studied radial grids
can be found in the SI.

Having established the computational methodology, we will proceed
to discuss functionals that show signs of numerical ill behavior.
However, this requires first answering the question posed in \secref{Theory}
of what constitutes ``quick enough'' convergence, as this is the
criterion used in this work to determine numerical ill behavior in
density functionals. In fact, quite a bit of work has been dedicated
to answering this question in the literature (as well as in determining
the defaults of various quantum chemistry programs) in the case of
well-behaved functionals.

For example, the standard grids (SG) originally developed by \citeauthor{Gill1993_CPL_506}
contain up to 26 radial points per atom for SG-0,\citep{Chien2006_JCC_730}
50 for SG-1,\citep{Gill1993_CPL_506} 75 for SG-2,\citep{Dasgupta2017_JCC_869}
and 99 for SG-3.\citep{Dasgupta2017_JCC_869} SG-1 is well known to
yield sufficiently converged energies for LDAs and most GGAs; SG-2
is recommended for tougher GGAs and most meta-GGAs, while SG-3 is
recommended for Minnesota functionals.\citep{Dasgupta2017_JCC_869}
Although the used number of radial grid points may depend on the atom,
many other quantum chemistry programs also use around 100 radial quadrature
points in their default grids. 

Such quadratures clearly are sufficient for reaching $\mu E_{h}$
level accuracy with well-behaving density functionals exemplified
in \figref{TPSS-exchange}. The maximal quadrature errors for TPSS
exchange shown in \figref{TPSS-exchange} are $1.527\times10^{-4}E_{h}$
for 50 radial points (similarly to SG-1), $3.496\times10^{-6}E_{h}$
for 75 radial points (similarly to SG-2, which would be the recommended
default grid for TPSS exchange), and $5.188\times10^{-8}E_{h}$ for
100 radial points (1 more than in SG-3). However, we will demonstrate
that many functionals require way more radial quadrature points to
achieve sub-$\mu E_{h}$ converged total energies.

The following discussion will thereby focus on functionals that do
not behave as the functionals exemplified by \figref{TPSS-exchange}:
ones that either require hundreds more grid points to converge to
machine precision, and ones that fail to converge to machine precision
even with unreasonably many (2500) radial quadrature points.

\subsection{Ill-behaved LDAs \label{subsec:Ill-behaved-LDAs}}

The \citeyearpar{GellMann1957_PR_364} functional by \citet{GellMann1957_PR_364}
(Libxc identifier LDA\_C\_RPA), $\epsilon_{c}=a\log r_{s}+b+cr_{s}\log r_{s}+dr_{s}$,
where $r_{s}$ is the Wigner--Seitz radius of the electron density,
appears grid sensitive. The Ahlrichs M4 grid (\schemeref{Ahlrichs-M4})
affords quick and robust convergence, whereas the Ahlrichs M3 grid
(\schemeref{Ahlrichs-M3}) shows long tails. The Mura--Knowles scheme
yields poor results for this functional.

The \citeyearpar{Gordon1972_JCP_3122} correlation functional by \citet{Gordon1972_JCP_3122}
(LDA\_C\_GK72) has a piecewise definition, which is likely the cause
for the poor convergence behavior shown in the SI; the quadrature
error saturates to $|\Delta E|=\mathcal{O}(10^{-4}E_{h})$.

The third \citeyearpar{Vosko1980_CJP_1200} correlation functional
by \citet{Vosko1980_CJP_1200} (LDA\_C\_VWN\_3; defined in their equation
4.7) contains a ratio that is tentatively the reason for the observed
odd behavior, which is characterized by sharp features and rapid oscillations
with long-range order. Quadrature errors up to $|\Delta E|=\mathcal{O}(10^{-6}E_{h})$
are observed for the N and P atoms for this functional.

Importantly, the functional form recommended by \citeauthor{Vosko1980_CJP_1200},
usually known as VWN (available in Libxc as LDA\_C\_VWN, and sometimes
also known as VWN5) is well-behaved. Other VWN variants are well-behaved
as well. Importantly, this includes the version which is used in the
B3LYP functional\citep{Stephens1994_JPC_11623} that is based on random
phase approximation data (LDA\_C\_VWN\_RPA) instead of the more accurate
quantum Monte Carlo data used in VWN5; unfortunately, this version
is called VWN in the Gaussian program.\citep{Hertwig1997_CPL_345}

The \citeyearpar{Perdew1981_PRB_5048} correlation functional by \citet{Perdew1981_PRB_5048}
(LDA\_C\_PZ) is widely available in various electronic structure programs.
The functional has a piecewise definition with a cusp, leading to
sketchy convergence with numerical noise in the order $|\Delta E|=\mathcal{O}(10^{-6}E_{h})$
persisting even with thousands of grid points. The \citeyearpar{Ortiz1994_PRB_1391}
correlation functional by \citet{Ortiz1994_PRB_1391,Ortiz1997_PRB_9970}
(LDA\_C\_OB\_PZ) shares the form of the PZ functional, likewise leading
to an apparent lack of convergence, while the parametrization of \citet{Ortiz1994_PRB_1391,Ortiz1997_PRB_9970}
of the Perdew--Wang functional (LDA\_C\_OB\_PW) is well-behaved.

The \citeyearpar{Liu1996_PRA_2211} correlation functional by \citet{Liu1996_PRA_2211}
(LDA\_C\_LP96) has a simple form
\[
\epsilon_{\text{xc}}(n)=C_{1}+C_{2}n^{-1/3}+C_{3}n^{-2/3},
\]
yet the convergence is very slow, and sharp features in the plot are
observed for the Li and Na atoms. Also this functional appears to
plateau to a quadrature error around $|\Delta E|=\mathcal{O}(10^{-6}E_{h})$.
The kinetic energy functional defined in the same paper (LDA\_K\_LP96)
employs the same functional form, and similarly shows poor convergence. 

The \citeyearpar{Proynov2009_PRA_14103} correlation functional of
\citet{Proynov2009_PRA_14103} (LDA\_C\_PK09) is found to be ill-behaved.
However, the ill behavior may be caused by differences in the thresholding
of the various denominators appearing in the functional's equations
that appear to be used in the authors' reference implementation, but
have not been described in \citeref{Proynov2009_PRA_14103}.

\subsection{Ill-behaved GGAs \label{subsec:Ill-behaved-GGAs}}

The \citeyearpar{Herman1969_PRL_807} exchange functional by \citeauthor{Herman1969_PRL_807}\citep{Herman1969_PRL_807,Herman1970_IJQC_827}
(GGA\_X\_HERMAN) has a simple functional form---the enhancement factor
is $F_{x}(x)=1+cx^{2}$---yet there is noticeable noise in the energy;
the quadrature error plateaus quickly to $|\Delta E|=\mathcal{O}(10^{-8}E_{h})$.
This is not suprising, as the enhancement factor diverges in the asymptotic
limit as $x\rightarrow\infty$.

The \citeyearpar{Meyer1976_ZNA_898} kinetic energy functional by
\citet{Meyer1976_ZNA_898} (GGA\_K\_MEYER) is likewise ill-behaved.
The functional form includes a logarithm of a quantity that has a
denominator that can diverge; thus the stability of the quadrature
observed---plateauing to $|\Delta E|=\mathcal{O}(10^{-5}E_{h})$---is
surprisingly good. 

The \citeyearpar{Langreth1981_PRL_446} correlation functional by
\citet{Langreth1981_PRL_446} (GGA\_C\_LM) was one of the first GGA
functionals, and its convergence leaves something to be desired: the
quadrature error plateaus to $|\Delta E|=\mathcal{O}(10^{-7}E_{h})$.

The \citeyearpar{Perdew1986_PRB_8822} correlation functional by \citet{Perdew1986_PRB_8822}
(GGA\_C\_P86) was one of the first successful GGA correlation functionals.
It is based on the PZ LDA, and inherits its poor convergence behavior,
plateauing to $|\Delta E|=\mathcal{O}(10^{-6}E_{h})$. Interestingly,
this functional has been also used in some fully numerical studies.\citep{Brakestad2021_JCP_214302}
A variant based on VWN (that is, VWN5) correlation is available in
several programs, and this variant (GGA\_C\_P86VWN) is numerically
well-behaved.

The \citeyearpar{Lacks1993_PRA_4681} exchange functional by \citet{Lacks1993_PRA_4681}
(GGA\_X\_LG93) has a complicated form with a high-order polynomial,
a root and a well-behaved denominator. As a result, the functional
requires several hundred quadrature points to converge to machine
precision.

The \citeyearpar{Filatov1997_MP_847} correlation functional by \citealt{Filatov1997_IJQC_603,Filatov1997_MP_847}
(GGA\_C\_FT97) is characterized by slow convergence. Over 1000 quadrature
points are required to reduce the quadrature error to the plateaued
value $|\Delta E|=\mathcal{O}(10^{-11}E_{h})$.

The range separated $\omega$PBEh functional by \citet{Ernzerhof1998_JCP_3313,Heyd2003_JCP_8207,Heyd2004_JCP_7274,Heyd2006_JCP_219906,Henderson2009_JCP_44108}
(GGA\_X\_WPEH) is characterized by a surprising amount of numerical
noise, quickly plateauing to an error around $|\Delta E|=\mathcal{O}(10^{-5}E_{h})$.
This behavior is also carried out into the range separated hybrids
that are based on the $\omega$PBEh functional, such as the HSE03\citep{Heyd2003_JCP_8207}
and HSE06\citep{Heyd2006_JCP_219906} functionals.

The \citeyearpar{Gilbert1999_CPL_511} exchange functional by \citet{Gilbert1999_CPL_511}
(GGA\_X\_GG99) appears to yield energies that are susceptible to numerical
noise, as the quadrature error plateaus to $|\Delta E|=\mathcal{O}(10^{-6}E_{h})$.

The \citeyearpar{Tsuneda1999_JCP_10664} one-parameter progressive
correlation functional by \citet{Tsuneda1999_JCP_10664,Tsuneda1999_JCP_5656}
based on the PW91 exchange functional\citep{Perdew1992_PRB_6671,Perdew1993_PRB_4978}
(GGA\_C\_OP\_PW91) is extremely ill-behaved, plateauing to $|\Delta E|=\mathcal{O}(1\ E_{h})$
for Li and Na. The variants based on other exchange functionals are
better behaved.

The \citeyearpar{Henderson2008_JCP_194105} range-separated exchange
functional by \citet{Henderson2008_JCP_194105} based on the \citeyearpar{Becke1988_PRA_3098}
exchange functional of \citet{Becke1988_PRA_3098} (GGA\_X\_HJS\_B88)
appears noisy, plateauing to $|\Delta E|=\mathcal{O}(10^{-4}E_{h})$.
Also the later version introduced in \citeyearpar{Weintraub2009_JCTC_754}
by \citet{Weintraub2009_JCTC_754} (GGA\_X\_HJS\_B88\_V2) does not
appear to be smooth, although it plateaus to a smaller quadrature
error of $|\Delta E|=\mathcal{O}(10^{-6}E_{h})$. The functionals
of \citet{Henderson2008_JCP_194105} based on other exchange functionals
(GGA\_X\_HJS\_PBE, GGA\_X\_HJS\_PBE\_SOL, and GGA\_X\_HJS\_B97X) appear
to be smoother, but still require several hundred grid points to converge
to machine precision.

The \citeyearpar{Haas2011_PRB_205117} exchange functional by \citet{Haas2011_PRB_205117}
(GGA\_X\_HTBS) is a limited-range spline interpolation between the
GGA exchange functionals of \citet{Hammer1999_PRB_7413} (GGA\_X\_RPBE)
and \citet{Wu2006_PRB_235116} (GGA\_X\_WC). Although the quadratures
for the latter two converge rapidly, the limited-range spline interpolation
makes the HTBS enhancement function less smooth and quadrature errors
around $|\Delta E|=\mathcal{O}(10^{-10}E_{h})$ persist even with
over 1000 radial quadrature points. (The recent CASE21 machine learned
GGA by \citet{Sparrow2022_JPCL_6896}, HYB\_GGA\_XC\_CASE21, shows
similar convergence.)

The \citeyearpar{Wellendorff2012_PRB_235149} exchange functional
by \citet{Wellendorff2012_PRB_235149} (GGA\_X\_BEEFVDW) functional
is somewhat grid sensitive, and plateaus to a level of numerical noise
of $|\Delta E|=\mathcal{O}(10^{-10}E_{h})$.

The \citeyearpar{Fabiano2014_JCTC_2016} correlation functional by
\citet{Fabiano2014_JCTC_2016} (GGA\_C\_GAPLOC) exhibits remarkably
slow convergence for a GGA functional, requiring around 600 radial
points for Ar to reach machine precision. The \citeyearpar{Margraf2019_JCP_244116}
correlation functional by \citet{Margraf2019_JCP_244116} (GGA\_C\_CCDF)
also exhibits a similar issue, requiring some 800 radial points for
Ar to reach machine precision.

\subsection{Ill-behaved meta-GGAs \label{subsec:Ill-behaved-meta-GGAs}}

The \citeyearpar{Becke1994_IJQC_625} correlation functional by \citet{Becke1994_IJQC_625}
(MGGA\_C\_B94) appears well-behaved but requires hundreds of radial
quadrature points for reliable convergence; around 600 are needed
to converge N to machine precision. The \citeyearpar{Becke1998_JCP_2092}
exchange-correlation functional by \citet{Becke1998_JCP_2092} (MGGA\_C\_B98)
has a similar behavior, likewise requiring around 700 radial quadrature
points to reach machine precision.

The \citeyearpar{Jemmer1995_PRA_3571} exchange functional by \citet{Jemmer1995_PRA_3571}
(MGGA\_X\_JK) is the poster child of numerically unstable functionals.
This meta-GGA contains a denominator that can vanish, leading to a
significant grid dependence in the functional. The functional was
characterized non-self-consistently in \citeref{Jemmer1995_PRA_3571},
and yields gigantic quadrature errors $|\Delta E|\ge\mathcal{O}(1\ E_{h})$
in our study.

The \citeyearpar{Filatov1998_PRA_189} exchange functional by \citet{Filatov1998_PRA_189}
(MGGA\_X\_FT98) is pronouncedly ill-behaved, exhibiting quadrature
errors around $|\Delta E|=\mathcal{O}(10^{-5}E_{h})$ even with thousands
of grid points.

The \citeyearpar{Rey1998_IJQC_581} correlation functional by \citet{Rey1998_IJQC_581,Kurth1999_IJQC_889,Krieger2001_ACP_48,Toulouse2002_JCP_10465}
(MGGA\_C\_KCISK) appears to be susceptible to numerical noise, quickly
plateauing to $|\Delta E|=\mathcal{O}(10^{-6}E_{h})$.

\begin{figure}
\begin{centering}
\includegraphics[width=0.5\textwidth]{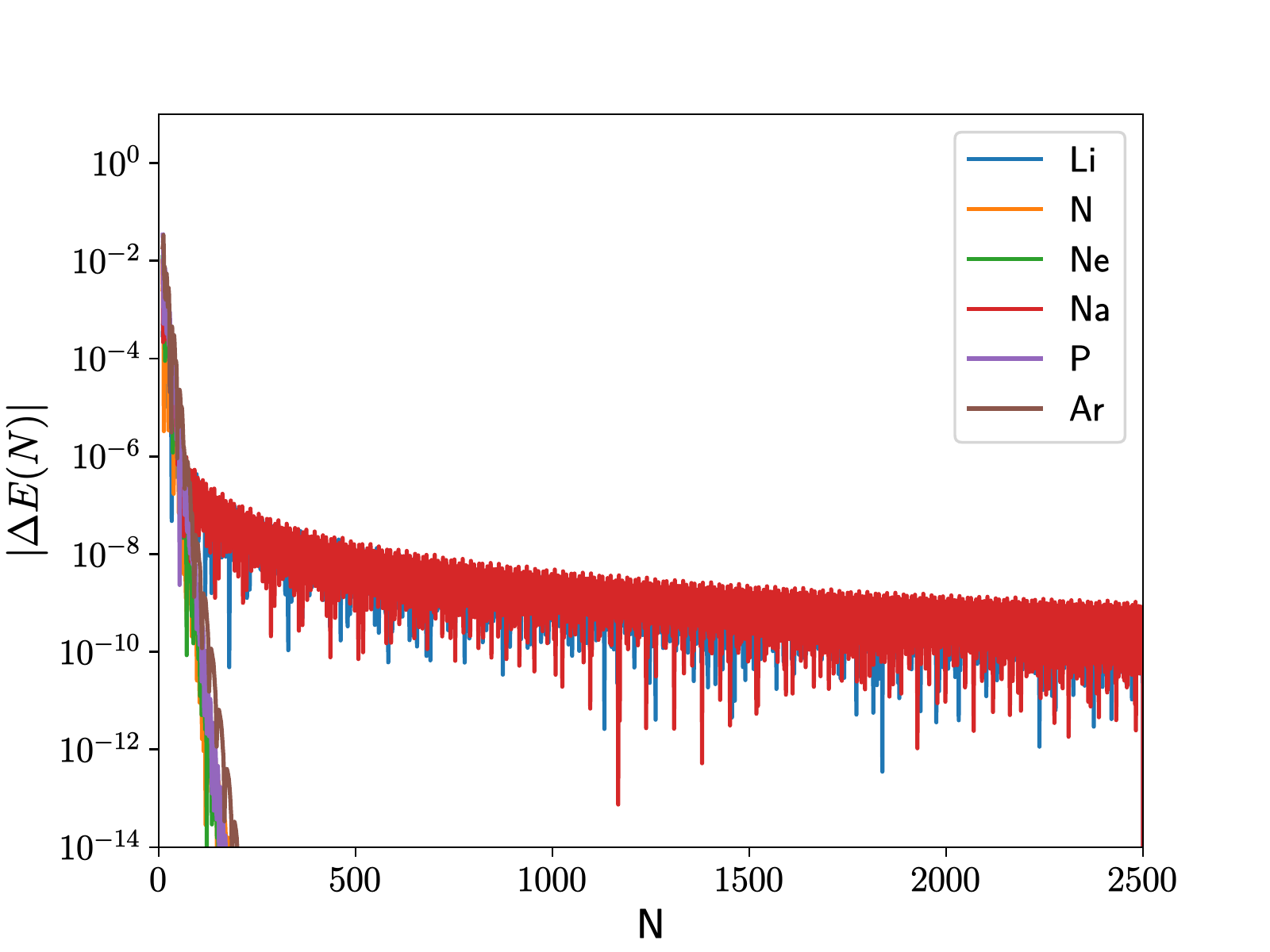}
\par\end{centering}
\caption{Quadrature error (\eqref{quad-error}) for the TPSS meta-GGA correlation
functional as a function of the number of radial points. \label{fig:TPSS-meta-GGA-correlation}}
\end{figure}

The \citeyearpar{Tao2003_PRL_146401} correlation functional by \citet{Tao2003_PRL_146401,Perdew2004_JCP_6898}
(MGGA\_C\_TPSS) plateaus to an error $|\Delta E|=\mathcal{O}(10^{-8}E_{h})$
for alkali atoms, as shown in \figref{TPSS-meta-GGA-correlation}.
Similar results are also observed for related functionals, including
the functional of \citet{Constantin2012_PRB_35130} (MGGA\_C\_TPSSLOC);
the revised correlation functional by \citet{Perdew2009_PRL_26403,Perdew2011_PRL_179902}
(MGGA\_C\_REVTPSS); the correlation functional by \citet{Tao2016_PRL_73001}
(MGGA\_C\_TM); and the correlation functional of \citet{Jana2019_JPCA_6356}
(MGGA\_C\_REVTM).

The \citeyearpar{Cancio2006_PRB_81202} exchange-correlation functional
of \citet{Cancio2006_PRB_81202} (MGGA\_XC\_CC06) is numerically unstable
like the \citet{Jemmer1995_PRA_3571} functional, as it too has a
denominator that can vanish. This is the likely cause for the observed
lack of convergence and residual quadrature errors that can reach
$|\Delta E|\ge\mathcal{O}(1\ E_{h})$ even with thousands of radial
grid points.

Given several hundred grid points, the \citeyearpar{Becke1989_PRA_3761}
exchange functional of \citet{Becke1989_PRA_3761} (MGGA\_X\_BR89)
plateaus to a quadrature error $|\Delta E|<\mathcal{O}(10^{-10}E_{h})$.
However, the \citeyearpar{Proynov2008_CPL_103} refit by \citet{Proynov2008_CPL_103}
(MGGA\_X\_BR89\_EXPLICIT) appears to increase numerical noise by three
orders of magnitude; resulting in a plateauing to $|\Delta E|=\mathcal{O}(10^{-7}E_{h})$.

The \citeyearpar{Loos2017_JCP_114108} exchange functionals of \citet{Loos2017_JCP_114108}
(MGGA\_X\_GX and MGGA\_X\_PBE\_GX) appear to be extremely ill-behaved,
slowly plateauing to $|\Delta E|\approx\mathcal{O}(10^{-4}E_{h})$
for Ar. Both functionals include a step function in the definition,
which may be the origin of the poor convergence.

The \citeyearpar{Patra2019_PRB_155140} exchange functional of \citet{Patra2019_PRB_155140}
(MGGA\_X\_MGGAC) requires several hundred radial grid points to achieve
converged total energies. Around 500 radial points are required for
the N atom to reach the plateau at $|\Delta E|=\mathcal{O}(10^{-12}E_{h})$.
Slightly fewer quadrature points are necessary for the \citeyearpar{Jana2021_JCP_24103}
exchange functional by \citet{Patra2020_JCP_184112} (MGGA\_X\_REGTM),
which reaches machine precision for all studied atoms with fewer than
500 radial points. However, the \citeyearpar{Jana2021_JCP_24103}
correlation functional (MGGA\_C\_RREGTM) by \citet{Jana2021_JCP_24103}
shows pathologically slow convergence to the grid limit, with an error
of the order $|\Delta E|=\mathcal{O}(10^{-8}E_{h})$ with 1000 radial
quadrature points, $|\Delta E|=\mathcal{O}(10^{-10}E_{h})$ with 2000
quadrature points. 

The \citeyearpar{Patra2019_PRB_45147} exchange functionals of \citet{Patra2019_PRB_45147}
(MGGA\_X\_MBRXC\_BG and MGGA\_X\_MBRXH\_BG) quickly stagnate to $|\Delta E|=\mathcal{O}(10^{-7}E_{h})$
precision.

The \citeyearpar{Brown2021_JCC_2004} machine learned density functional
of \citet{Brown2021_JCC_2004} (MGGA\_X\_MCML) requires hundreds of
radial grid points for accurate energies. Around 600 radial points
are necessary to converge all atoms to machine precision.

\subsection{Kinetic meta-GGAs}

The dependence on the Laplacian of the density in kinetic energy meta-GGAs
makes the functionals less well behaved. This is demonstrated by the
\citeyearpar{Hodges1973_CJP_1428} gradient expansions of \citet{Hodges1973_CJP_1428}
to the second (MGGA\_K\_GEA2) and fourth (MGGA\_K\_GEA4) order, the
first of which is well-behaved and plateaus to an error of $|\Delta E|=\mathcal{O}(10^{-12}E_{h})$,
while the latter is ill-behaved and plateaus at $|\Delta E|=\mathcal{O}(10^{-5}E_{h})$.

The convergence for the \citeyearpar{Perdew2007_PRB_155109} kinetic
energy functional of \citet{Perdew2007_PRB_155109} (MGGA\_K\_PC07)
is slow and depends on the parameters used. With the original parameters,
slow convergence to $|\Delta E|=\mathcal{O}(10^{-8}E_{h})$ for N
with $\mathcal{O}(2500)$ grid points is observed. With the parameters
reoptimized by \citet{MejiaRodriguez2017_PRA_52512} for deorbitalization
of the SCAN family (MGGA\_K\_PC07\_OPT), the functional plateaus to
a quadrature error of $|\Delta E|=\mathcal{O}(10^{-5}E_{h})$ for
Ar.

The \citeyearpar{Karasiev2009_PRB_245120} reduced derivative approximation
(MGGA\_K\_RDA) kinetic energy functional of \citet{Karasiev2009_PRB_245120}
is similarly characterized by slow convergence. Sharp oscillatory
features are observed for Ar, and errors around $|\Delta E|=\mathcal{O}(10^{-7}E_{h})$
are still observed with $\mathcal{O}(2500)$ grid points.

The \citeyearpar{Cancio2016_JCP_84107} CSK1, CSK4, CSK-LOC1, and
CSK-LOC4 kinetic energy functionals of \citet{Cancio2016_JCP_84107}
(MGGA\_K\_CSK1, MGGA\_K\_CSK4, MGGA\_K\_CSK\_LOC1, and MGGA\_K\_CSK\_LOC4,
respectively) converge slowly, with the slowest convergence observed
for nitrogen. More than 500 radial points are necessary to converge
N to $\mu E_{h}$ accuracy, and quadrature errors of $|\Delta E|=\mathcal{O}(10^{-10}E_{h})$
remain with $\mathcal{O}(2500)$ grid points with CSK4 and CSK-LOC4
showing rapid oscillations.

The \citeyearpar{Constantin2018_JPCL_4385} semilocal Pauli--Gaussian
(XC\_MGGA\_K\_PGSL025) kinetic energy functional of \citet{Constantin2018_JPCL_4385}
appears quite ill-behaved, quickly plateauing to $|\Delta E|=\mathcal{O}(10^{-4}E_{h})$.

\subsection{The SCAN family}

The most interesting examples of ill-behaved functionals are the recent
meta-GGAs constructed from first principles: although the TPSS functional
was found to be well-behaved, its successors are not. The MS0\citep{Sun2012_JCP_51101}
(MGGA\_X\_MS0) and MS2\citep{Sun2013_JCP_44113} (MGGA\_X\_MS2) functionals
require about twice the number of quadrature points to reach machine
precision compared to TPSS: 500 and 550, respectively, with the more
recent MS2 requiring more points than MS0. 

The successor to MS0 and MS2 is the MVS functional (MGGA\_X\_MVS),\citep{Sun2015_PNASUSA_685}
which again roughly doubles the required number of quadrature points:
most of the studied atoms reach machine precision with 1000 radial
quadrature points, while N requires about 1300 radial quadrature points.

\begin{figure}
\begin{centering}
\includegraphics[width=0.5\textwidth]{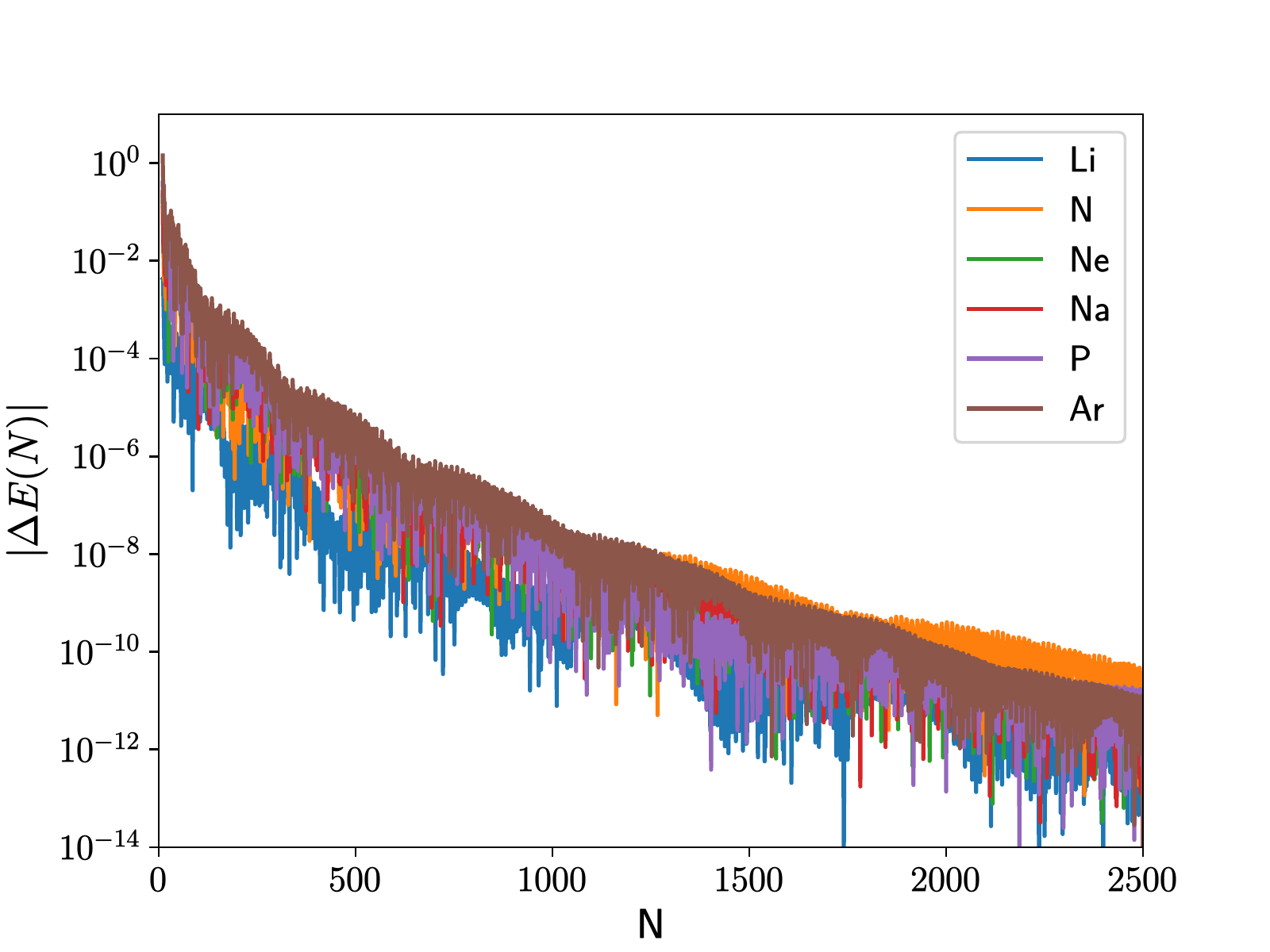}
\par\end{centering}
\caption{Quadrature error (\eqref{quad-error}) for the SCAN meta-GGA exchange
functional as a function of the number of radial points. \label{fig:SCAN-meta-GGA-exchange}}
\end{figure}

Finally, the SCAN functional\citep{Sun2015_PRL_36402} (MGGA\_X\_SCAN,
shown in \figref{SCAN-meta-GGA-exchange})---which is well-known
to be numerically ill-behaved\citep{Bartok2019_JCP_161101}---is
clearly problematic as is shown by its remarkably slow convergence
rate. Around 600 radial quadrature points are required to reach microhartree
precision, and errors around $|\Delta E|=\mathcal{O}(10^{-10}E_{h})$
persist even with $\mathcal{O}(2500)$ radial quadrature points.

\begin{figure}
\begin{centering}
\includegraphics[width=0.5\textwidth]{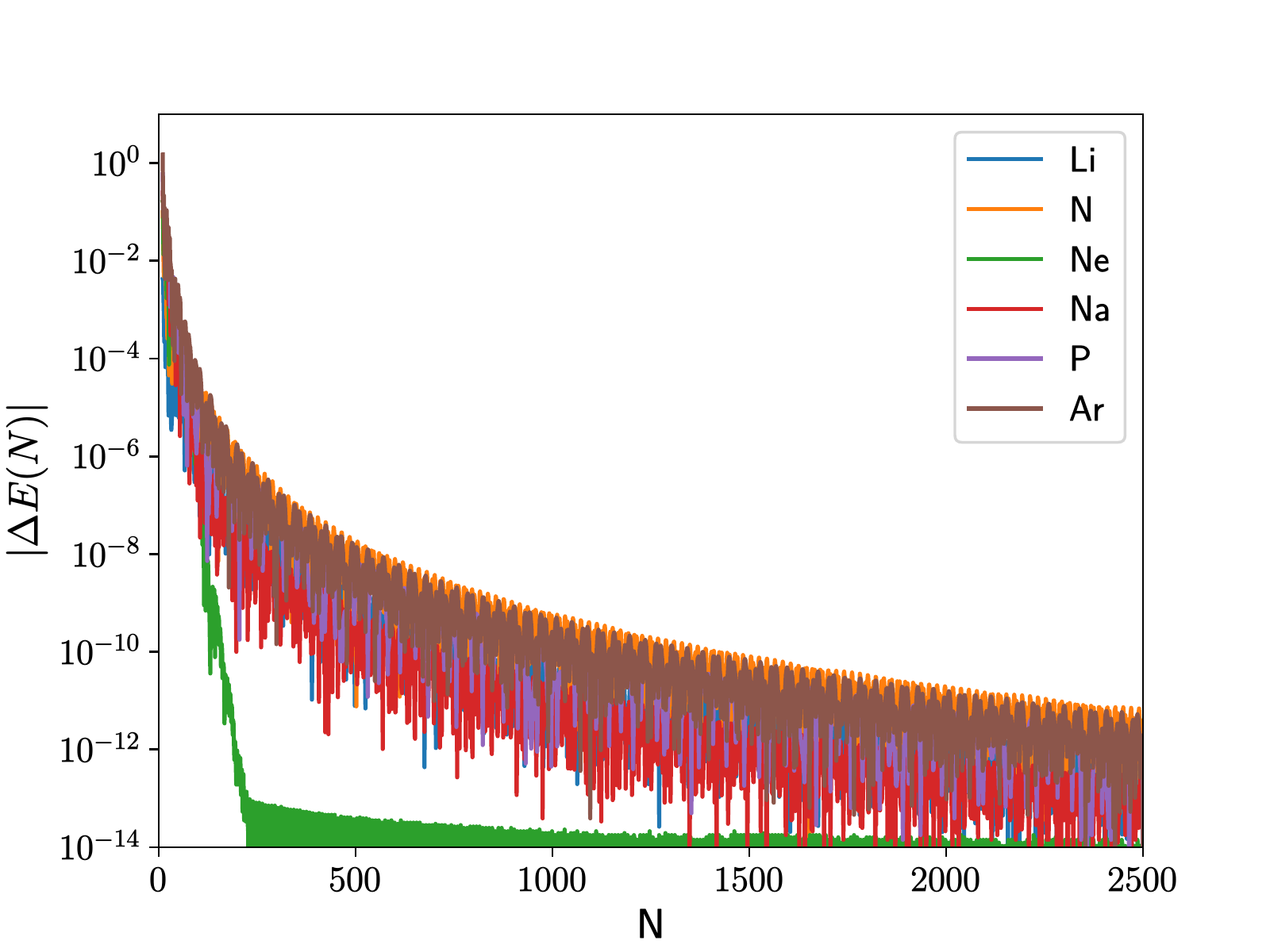}
\par\end{centering}
\caption{Quadrature error (\eqref{quad-error}) for the rSCAN meta-GGA exchange
functional as a function of the number of radial points. \label{fig:rSCAN-meta-GGA-exchange}}
\end{figure}

The regularized SCAN (rSCAN) functional (MGGA\_X\_RSCAN, shown in
\figref{rSCAN-meta-GGA-exchange}) was designed to fix the issues
with numerical behavior in SCAN.\citep{Bartok2019_JCP_161101} Although
the convergence is clearly improved for Ne, the functional is still
found to converge extremely slowly to the quadrature limit at fixed
density for the other atoms: a similar $|\Delta E|=\mathcal{O}(10^{-10}E_{h})$
level error is observed even with $\mathcal{O}(2500)$ radial quadrature
points.

\begin{figure}
\begin{centering}
\includegraphics[width=0.5\textwidth]{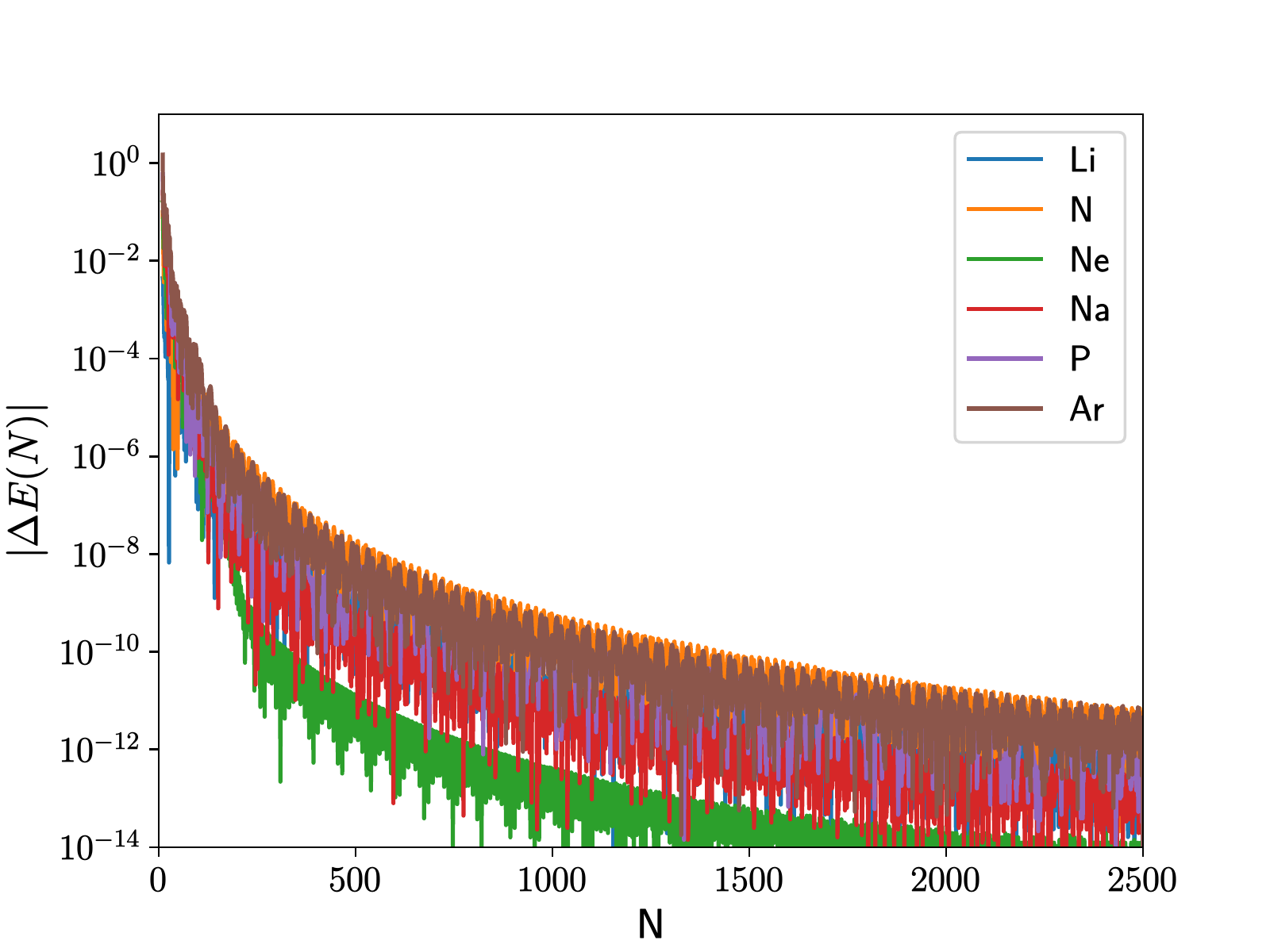}
\par\end{centering}
\caption{Quadrature error (\eqref{quad-error}) for the r$^{2}$SCAN meta-GGA
exchange functional as a function of the number of radial points.
\label{fig:r2SCAN-meta-GGA-exchange}}
\end{figure}

Recent functionals that aim to restore constraint adherence to rSCAN---the
r++SCAN and r$^{4}$SCAN functionals\citep{Furness2022_JCP_34109}
(MGGA\_X\_RPPSCAN and MGGA\_X\_R4SCAN, respectively) as well as the
r$^{2}$SCAN (MGGA\_X\_R2SCAN, shown in \figref{r2SCAN-meta-GGA-exchange})
functional\citep{Furness2020_JPCL_8208,Furness2020_JPCL_9248}---are
also found to behave similarly. The convergence for r++SCAN is strikingly
similar to rSCAN, with the main difference being that the good behavior
for Ne in rSCAN is lost in r++SCAN. r$^{4}$SCAN has distinctly different
behavior to rSCAN and r++SCAN at small numbers of quadrature points,
but shares the slow asymptotic behavior for large numbers of grid
points and the $|\Delta E|=\mathcal{O}(10^{-10}E_{h})$ error with
$\mathcal{O}(2500)$ radial quadrature points. The behavior of r$^{2}$SCAN
appears identical to that of r++SCAN.

\citet{Holzwarth2022_PRB_125144} have recently proposed a modification
of r$^{2}$SCAN to reduce its undesireable numerical instabilities
found in fully numerical calculations on atoms. The modification of
increasing the $\eta$ parameter from $\eta=0.001$ in r$^{2}$SCAN
to $\eta=0.01$ in the r$^{2}$SCAN01 functional\citep{Holzwarth2022_PRB_125144}
(MGGA\_X\_R2SCAN01) does not result in any better convergence; the
r$^{2}$SCAN01 plots appear similar to the r$^{2}$SCAN ones. Similar
observations can also be made about the corresponding correlation
functionals for the whole SCAN family.

Unsurprisingly, the issues also affect the deorbitalized functionals,\citep{MejiaRodriguez2017_PRA_52512,MejiaRodriguez2018_PRB_115161}
SCAN-L\citep{Sun2015_PRL_36402,MejiaRodriguez2017_PRA_52512,MejiaRodriguez2018_PRB_115161}
(MGGA\_X\_SCAN\_L) and r$^{2}$SCAN-L\citep{Furness2020_JPCL_8208,Furness2020_JPCL_9248,MejiaRodriguez2020_PRB_121109}
(MGGA\_X\_R2SCAN\_L). Noting that issues with numerical ill behavior
were also found above in the PC07 kinetic energy functional used for
the deorbitalization, it comes as no surprise that SCAN-L and r$^{2}$SCAN-L
are less well behaved than SCAN and r$^{2}$SCAN, respectively, and
plateau to errors around $|\Delta E|=\mathcal{O}(10^{-8}E_{h})$ even
with $\mathcal{O}(2500)$ radial quadrature points.

\begin{figure}
\begin{centering}
\includegraphics[width=0.5\textwidth]{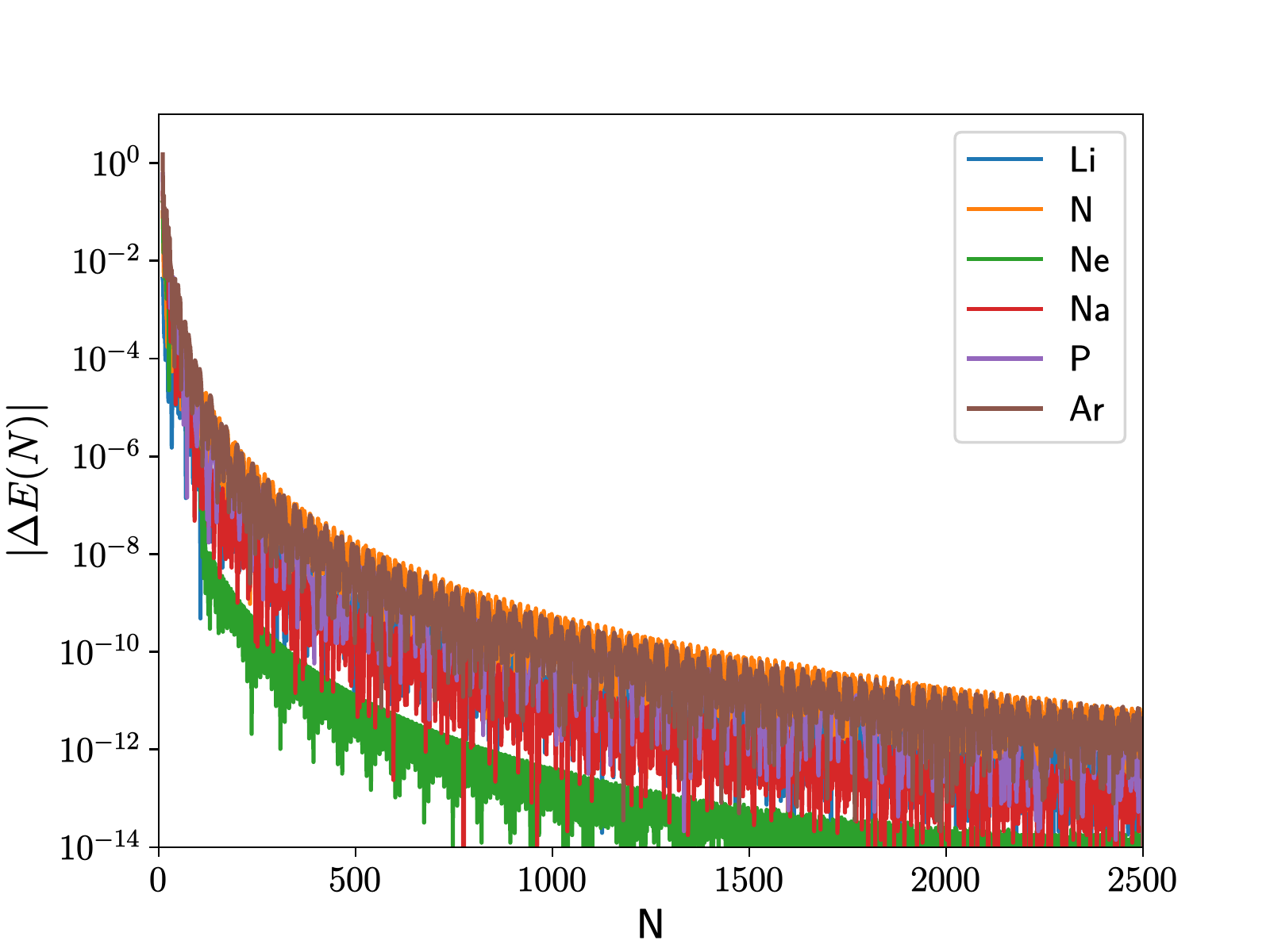}
\par\end{centering}
\caption{Quadrature error (\eqref{quad-error}) for the r++SCAN meta-GGA exchange
functional as a function of the number of radial points. \label{fig:r++SCAN-meta-GGA-exchange}}
\end{figure}

\section{Summary and Discussion \label{sec:Summary-and-Discussion}}

Being able to converge total energies to high precision is of utmost
importance for both the theory and applications of DFT. As pointed
out in \secref{Introduction}, this requires three things from a given
density functional:
\begin{enumerate}
\item the total energy must be evaluatable accurately at fixed density,
\label{enu:the-total-energy}
\item SCF calculations must be easily convergeable, and \label{enu:self-consistent-field-calculatio}
\item the complete basis set limit must be reachable. \label{enu:the-complete-basis}
\end{enumerate}
Any functional that breaks these criteria is deemed numerically ill-behaved.

Inaccuracies in total energy evaluation (criterion \ref{enu:the-total-energy})
may result in a lack of self-consistent field convergence, leading
to breakage of criterion \ref{enu:self-consistent-field-calculatio}
that can be a major issue for applications of DFT. Numerical well-behavedness
according to all the above criteria is especially important for fully
numerical approaches that target sub-$\mu E_{h}$ accurate total energies.
Likewise, the optimization of atomic basis sets for density functional
calculations requires numerically well-behaved functionals, as the
basis set parameters---such as the exponent in Gaussian or Slater
type orbital basis sets---are typically optimized to sub-$\mu E_{h}$
level precision in SCF calculations, even if the basis set truncation
error of the Gaussian basis set is larger than this.

We have studied the numerical behavior of all 592 density functionals
for three-dimensional systems included in Libxc 5.2.2 according to
criterion \ref{enu:the-total-energy} by employing tabulated Hartree--Fock
wave functions\citep{Koga1999_IJQC_491} for Li, N, Ne, Na, P, and
Ar. By examining the convergence of the quadrature of the density
functional energy at fixed density, we were able to demonstrate ill-behavedness
in a number of recent density functionals.

These results strongly suggest that issues with numerical behavior
of DFAs have not been adequately investigated nor addressed by the
density functional community. A practical functional needs to converge
rapidly to the grid limit for a fixed density with any reasonable
quadrature approach. Standard quadrature grids\citep{Gill1993_CPL_506,Chien2006_JCC_730,Dasgupta2017_JCC_869}
for density functional theory typically employ only $\mathcal{O}(100)$
radial quadrature points even for meta-GGA functionals. While such
grids are often suitable for well-behaved functionals, we have identified
a number of density functionals which do not converge to the level
of precision required in routine applications even with fixed electron
densities, or even if an unseemly large number of radial grid points
is used. Significant numerical issues were especially discovered in
the whole SCAN family of functionals, and we hope that future functionals
will be better behaved in this aspect.

One of the original motivations of this work is the reproducibility
of density functional approximations: how can we know if the implementation
of a given density functional is correct? The numerical behavior of
the density functional is an important aspect to consider in this
aspect, as reporting reference energies is highly desirable for enabling
reproducibility, as we will discuss in upcoming work. In short, reference
energies need to be evaluated accurately, at the quadrature limit,
in order to be fully reproducible. Also for this reason, the grid
sensitivity of the total energy should be considered as an essential
part in the development of new density functionals, as many functionals
studied in this work do not appear to allow the determination of sub-$\mu E_{h}$
accurate reference energies.

Although we have found several recent first principles meta-GGAs to
be numerically ill-behaved, many other recent meta-GGA functionals
show quick convergence to the quadrature limit. The key difference
between the two kinds of meta-GGA functionals appears to be that the
problematic physicists' first principles functionals include kinetic
energy dependence through the $\alpha$ parameter\textbf{ 
\begin{equation}
\alpha_{\sigma}=(\tau_{\sigma}-\tau_{\text{\ensuremath{\sigma}}}^{W})/\tau_{\sigma}^{\text{unif}},\label{eq:alpha}
\end{equation}
}where 
\begin{equation}
\tau_{\text{\ensuremath{\sigma}}}^{W}=|\nabla n_{\sigma}|^{2}/8n_{\sigma}\label{eq:tau-w}
\end{equation}
 and 
\begin{equation}
\tau_{\sigma}^{\text{unif}}=(3/10)(6\pi^{2})^{(2/3)}n_{\sigma}^{5/3}.\label{eq:tau-unif}
\end{equation}
This $\alpha$ parameter is closely related to the curvature of the
Fermi hole\citep{Dobson1991_JCP_4328} and to the factor \textbf{$D_{\sigma}=1-|\nabla n_{\sigma}|^{2}/(8n_{\sigma}\tau_{\sigma})$
}that leads to singularies that occur at critical points of the electron
density that are known to lead to instabilities.\citep{Graefenstein2007_JCP_214103}
In contrast, the chemists' functionals are based on a finite domain
transformation $-1\leq w_{\sigma}\leq1$ of the kinetic energy density
$\tau_{\sigma}$ as\citep{Zhao2008_JCTC_1849} 
\[
t_{\sigma}=\frac{\tau_{\sigma}^{\text{unif}}}{\tau_{\sigma}},\ \ w_{\sigma}=\frac{t_{\sigma}-1}{t_{\sigma}+1},
\]
which appears to lead to quickly convergent quadratures of the total
energy.

Another ingredient
\[
\beta_{\sigma}(\boldsymbol{r})=\alpha_{\sigma}(\boldsymbol{r})\frac{\tau_{\sigma}^{\text{unif}}(\boldsymbol{r})}{\tau_{\sigma}(\boldsymbol{r})+\tau_{\sigma}^{\text{unif}}(\boldsymbol{r})}
\]
was proposed by \citet{Furness2019_PRB_41119} and used in the MS2$\beta$
exchange functional, which is obtained from the numerically ill-behaved
MS2\citep{Sun2013_JCP_44113} exchange functional by replacing $\alpha$
with $2\beta$.\citep{Furness2019_PRB_41119} In our tests, MS2$\beta$
(MGGA\_X\_MS2B) appears numerically well-behaved, requiring only slightly
more quadrature points to converge to machine precision than the TPSS
functional, for instance. 

Finally, we wish to reiterate that any nasty behavior in total energy
evaluation that occurs at the presently studied fixed atomic densities
will also occur in self-consistent calculations. However, some functionals
that are well-behaved with respect to the energy may still turn out
to be unstable or slowly convergent in self-consistent calculations,
if the derivatives exhibit strong oscillations, large values, discontinuities,
singularities, and so on, and such numerical ill behavior has been
described in the literature for various meta-GGA functionals.\citep{Graefenstein2007_JCP_214103,Johnson2009_JCP_34111,Wheeler2010_JCTC_395,Mardirossian2013_JCTC_4453,Mardirossian2017_MP_2315,Bartok2019_JCP_161101,Lehtola2021_JCTC_943,Sitkiewicz2022_JPCL_5963}
These kinds of numerical issues---the breakage of criteria \ref{enu:self-consistent-field-calculatio}
and \ref{enu:the-complete-basis} above---can be probed with self-consistent
calculations in extended basis sets. Flexible fully numerical approaches
are arguably the strongest acid test for numerical behavior and they
will be described in another upcoming manuscript.

\section*{Conflict of Interest}

The authors have no conflicts to disclose.

\section*{Data Availability Statement}

The data that supports the findings of this study are available within
the supplementary material. 

\section*{Supporting Information}

Plots of the quadrature error for all studied density functionals
and radial grids.

\section*{Acknowledgments}

We thank James Furness for developing the original version of AtomicOrbitals,
which is a valuable tool for the community, as well as Ajit Thakkar
for supplying the atomic wave functions of \citerefs{Koga1999_IJQC_491}
and \citenum{Koga2000_TCA_411} in machine readable format. S.L. thanks
the National Science Foundation for financial support under grant
no. CHE-2136142, as well as the Academy of Finland for financial support
under project numbers 350282 and 353749.

\clearpage{}

\bibliography{citations}

\end{document}